\documentclass{article}
\usepackage{spconf,amsmath,graphicx}
\usepackage{multirow}
\usepackage{tabularx}
\usepackage{cite}
\usepackage{booktabs}
\usepackage{hyperref}
\hypersetup{colorlinks,allcolors=black}

\title{RescueSpeech: A German Corpus for Speech Recognition in Search and Rescue Domain}
\name{Sangeet Sagar$^{1,4}$, Mirco Ravanelli$^{2}$, Bernd Kiefer$^{1,4}$, Ivana Kruijff‑Korbayová$^4$, Josef van Genabith$^{1,4}$ }
\address{
  $^1$Saarland University, Germany\\
  $^2$Concordia University, Mila-Quebec AI Institute, Canada\\
  $^4$German Research Center for Artificial Intelligence (DFKI), Germany\\
\texttt{\small sangeetsagar2020@gmail.com}, \texttt{\small ravanellim@mila.quebec}, \\
\texttt{\small \{bernd.kiefer,josef.van\_genabith\}@dfki.de}, \texttt{\small ivana.kruijff@rettungsrobotik.de}\
  }

\begin{document}
%
\maketitle
\begin{abstract}
Despite the recent advancements in speech recognition, there are still difficulties in accurately transcribing conversational and emotional speech in noisy and reverberant acoustic environments. This poses a particular challenge in the search and rescue (SAR) domain, where transcribing conversations among rescue team members is crucial to support real-time decision-making.
The scarcity of speech data and associated background noise in SAR scenarios make it difficult to deploy robust speech recognition systems.

To address this issue, we have created and made publicly available a German speech dataset called \textit{RescueSpeech}. This dataset includes real speech recordings from simulated rescue exercises. Additionally, we have released competitive training recipes and pre-trained models. Our study highlights that the performance attained by state-of-the-art methods in this challenging scenario is still far from reaching an acceptable level.
\end{abstract}

\section{Introduction}
\label{sec:introduction}
Automatic speech recognition (ASR) can be crucial in situations like search and rescue (SAR) missions. These scenarios often involve making critical decisions in extremely hostile conditions, such as underground rescue operations, nuclear accidents, fire evacuation, or collapsed building after an earthquake. In such cases, rescue workers must act quickly and accurately to prevent the loss of lives and damage. Transcribing and automatically analyzing the conversations within the rescue team can provide useful support to help the team make the right decisions in a limited amount of time.
The context of search and rescue missions poses significant challenges for current speech recognition technologies. Speech recognizers must be able to handle conversational speech that is fast, emotional, and spoken under stressful conditions. Additionally, the acoustic environment in which rescuers operate is often extremely noisy, and recordings may be corrupted by various non-stationary noises, such as engine noise, vehicle sirens, radio chatter, helicopter noise, and other unpredictable disturbances. In recent years, there has been a significant amount of research focused on addressing these challenges \cite{willms2019,jmse8100818,ssar-2017}. Advanced deep learning techniques, such as self-supervised learning coupled with large datasets \cite{Mohamed_2022}, have been instrumental in achieving impressive performance improvements. One of the most intriguing aspects of the SAR domain is that all of the aforementioned challenges occur simultaneously, creating an incredibly difficult and complex task. This not only makes it an area of significant scientific interest but also underscores the urgent need for continued research and development in this field.

Developing a speech recognition system in this context is made even more challenging due to the limited availability of data in this critical domain. Collecting speech data related specifically to the SAR domain can be difficult, and privacy restrictions can often limit access to such data by the scientific community.
To encourage research in this field, we have released RescueSpeech\footnote{Available at: https://zenodo.org/record/8077622}, a German dataset for the Search and Rescue Domain Speech. This dataset contains authentic speech recordings between members of a rescue team during several rescue exercises. To the best of our knowledge, we are the first to publicly release an audio dataset in the SAR domain. RescueSpeech contains approximately 2 hours of annotated speech material. Although this amount may seem limited, it is actually quite valuable and can be effectively used to fine-tune large pretrained models such as wav2vec2.0 \cite{wav2vec2}, WavLM \cite{wavelm-2022}, and Whisper \cite{whisper-2022}. In fact, we demonstrate that this material is also suitable for training models from scratch when combined with proper data augmentation techniques and multi-condition training.

This paper presents a comprehensive collection of experimental evidence for the task at hand-- noise-robust German speech recognition. It employs state-of-the-art methods for both speech recognition and speech enhancement, as well as a combination of the two. Despite excelling in simpler scenarios, our results show that even modern ASR systems like Whisper \cite{whisper-2022}, struggle to perform well in the demanding rescue and search domain.
We have made our training recipes and pretrained models available to the community within the SpeechBrain toolkit
\footnote{Available at: \url{https://github.com/speechbrain/speechbrain/tree/develop/recipes/RescueSpeech}}
.
With the release of the RescueSpeech dataset we hope to foster research in this field and establish a common benchmark. We believe that our effort can help raise awareness about the importance of the use of speech technology in SAR missions, and the need for continued research in this domain.

\section{The RescueSpeech Dataset}
\label{sec:RescueSpeech_dataset}
RescueSpeech contains a blend of microphone and radio-recorded speech that includes excerpts from communication among robot-assisted emergency response team members during several simulated SAR exercises, which involves real firefighters speaking in high-stress situations like fire rescue, explosion etc. that can elicit heightened emotions. The speakers involved in the exercises are native speakers of the German language where conversations were carried out between team members, radio operators and the team leader. These dialogues loosely adopt a typical radio style communication wherein the start/end of a conversation is indicated by the use of certain words, connection quality is relayed, and acceptance or rejection of requests are conveyed. The practical use case of our dataset is limited not only for robot control but also for speech recognition, with its main application being the support of decision-makers and process monitors in disaster situations. The ASR output is analyzed by a natural language understanding (NLU) component and fused with sensor data, including GPS coordinates from robots or drones. This way we extract mission-related information from conversations and use it to offer assistance later in the deployment of the full system

\begin{figure}[t!]
  \centering
  \includegraphics[scale=0.7]{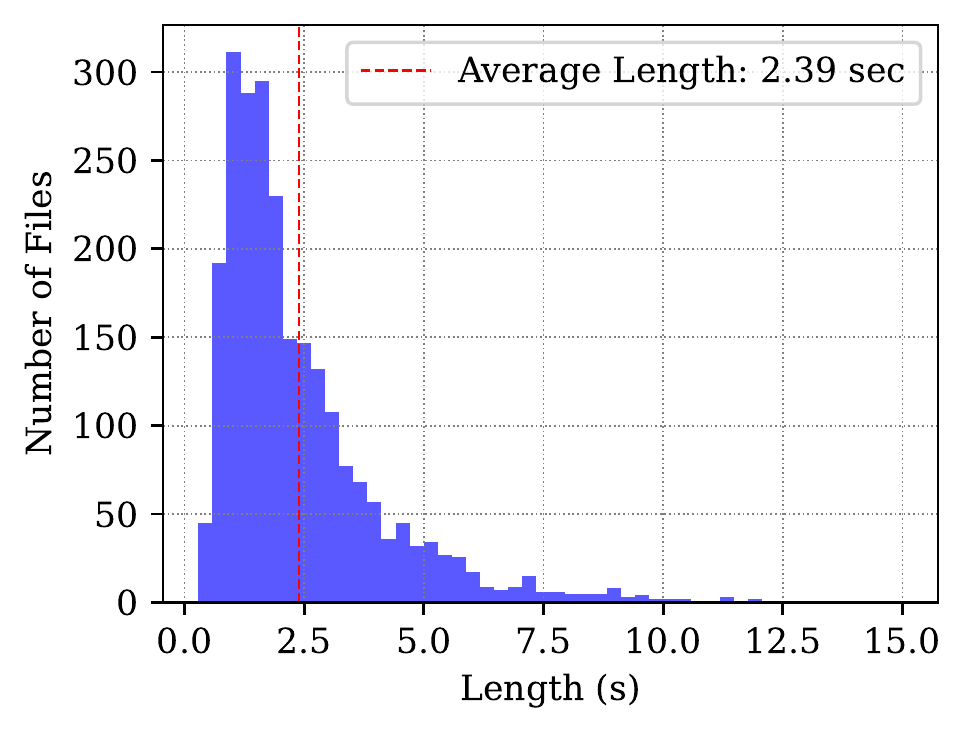}
  \caption{Histogram plot illustrating the average length of utterances in RescueSpeech in secs.}
  \label{fig:RescueSpeech_distribution}
\end{figure}

Initially captured at 44.1 kHz sampling rate, these recordings are down-sampled to 16 kHz, and further segmented to obtain a set of mono-speaker single-channel audio recordings. All utterances are also manually transcribed. The total length of the dataset is 1.6h with a total of 2412 sentences with 1591/245/576 sentences in train/valid/test set. We call it the RescueSpeech clean dataset. Figure \ref{fig:RescueSpeech_distribution} shows a histogram plot of the average length of the segmented utterances with an average length of 2.39 sec.
We also created a noisy version of RescueSpeech by contaminating our dataset with noisy clips from the AudioSet dataset \cite{audioset-2017} that includes five noise types-- \textit{emergency vehicle siren}, \textit{breathing}, \textit{engine}, \textit{chopper}, and \textit{static radio noise}. We utilized both real and synthetic room-impulse responses (RIR) (SLR26, SLR27 \cite{slr26}) to add reverberation as well. We then added noisy sequences to generate noisy utterances with different signal-to-noise ratios (SNR) (from -5 dB to 15 dB with a step of 1 dB). Each clean utterance is randomly corrupted with one of the noise types to generate 4500/1350/1350 train/valid/test utterances. We also ensure that a noise utterance used in the train set is only in this set. This randomness and exclusivity ensure that each split has an equal proportion for each noise type and that noises in each of the splits are different. This dataset provides a diverse set of noise and reverberation conditions that enable fine-tuning of our speech-enhancement model for improved accuracy on noisy RescueSpeech. We call this the RescueSpeech noisy dataset. Table \ref{tab:data_stat} briefly shows the distribution of utterances and duration for the clean and noisy version of the dataset.

\subsection{Related Corpora}
\label{ssec:related_corpora}
To improve the accuracy of speech recognition systems in noisy and reverberant environments, several corpora have been developed, such as CHIME \cite{chime3, chime4,chime5,chime6}, DIRHA \cite{dirha1,dirha2,dirha3,dirha4}, AMI \cite{ami}, VOiCES \cite{voices}, and COSINE \cite{cosine}. Among these, CHIME5 \cite{chime5} and CHIME6 \cite{chime6} are especially challenging because it contains conversational speech recorded during a dinner party in a domestic setting, where noise and reverberations are common. RescueSpeech also contains conversational speech recorded in challenging acoustic environments, but the scenario addressed in this corpus is unique and different from a dinner party. The acoustic conditions, emotions, and lexicon used in RescueSpeech are distinct, and thus provide an additional set of challenges for speech recognition systems.

The noisy version of RescueSpeech can be utilized to train speech enhancement systems that are robust in the acoustic conditions present in the Search and Rescue (SAR) domain. There are numerous datasets that have been released for speech enhancement purposes, including the deep-noise suppression (DNS) dataset \cite{dns22}, VoiceBank-DEMAND corpus \cite{voicebank}, and WHAM! and WHAMR! corpora  \cite{Wichern2019WHAM}, all of which are helpful for training speech enhancement models. However, the key difference with RescueSpeech is that it has been specifically designed for the SAR domain, where characteristic sounds such as sirens, radio signals, helicopters, trucks, and others affect the recordings. This unique characteristic of RescueSpeech makes it an especially valuable resource for training speech enhancement systems that can perform well in SAR environments.
\begin{table}[t!]
  \caption{Distribution of utterances and hours in the RescueSpeech clean and noisy dataset.}
  \label{tab:data_stat}
  \centering
  \begin{tabular}[th]{l  c  c |  c c}
    \toprule
    & \multicolumn{2}{c}{\textbf{Clean}} & \multicolumn{2}{c}{\textbf{Noisy}} \\
    \cmidrule{2-3}
    \cmidrule{4-5}
                            &   Mins  &   \#Utts.    &   HRS  &   \#Utts. \\
    \midrule
                    Train	&	61.86	&	1591	&	7.20	&	4500 \\
                    Valid	&	9.61	&	245	    &	2.16	&	1350 \\
                    Test	&	24.68	&	576	    &	2.16	&	1350  \\
    \bottomrule
  \end{tabular}
\end{table}

\section{Exerimental Setup}
\label{sec:experimental_setup}
We explored multiple training strategies to perform noise robust speech recognition. Speech recognizers and enhancement models are trained on large corpora and then fine-tuned and evaluated on RescueSpeech data.

\subsection{ASR training}
\label{ssec:asr}
We follow two approaches for ASR training: one based on sequence-to-sequence modeling (seq2seq) and another one based on the connectionist temporal classification (CTC) method. For the seq2seq model, we employ a CRDNN (convolutional, recurrent, and dense-neural network) architecture \cite{crdnn2015, crdnn2021}. The CRDNN encoder is trained on the full 1200h of the German CommonVoice corpus \cite{commonvoice}. Decoding uses an attentional-GRU decoder and a beam search coupled with an RNN-based language model (LM). The LM is trained on Tuda-De\footnote[2]{https://www.inf.uni-hamburg.de/en/inst/ab/lt/resources/data/acoustic-models.html} \cite{tuda-de} (8M sents), Leipzig news corpus \cite{goldhahn-etal-2012-building} (9M sents), and train transcripts of the CommonVoice corpus. For the CTC based models, we use wav2vec2.0, and WavLM  architecture as encoders for the ASR pipeline. These encoders use self-supervised approach for learning high-level contextualized speech representation. It needs no language model and decoding is performed using greedy search. For wav2vec2.0 and WavLM we use pre-trained encoders \texttt{\scriptsize{facebook/wav2vec2-large-xlsr-53-german}}\footnote[3]{https://huggingface.co/facebook/wav2vec2-large-xlsr-53-german} and
\texttt{\scriptsize{microsoft/wavlm-large}}\footnote[4]{https://huggingface.co/microsoft/wavlm-large} respectively. Additionally, we also employ the pre-trained Whisper \cite{whisper-2022} model \texttt{\scriptsize{openai/whisper\\-large-v2}}\footnote[5]{https://huggingface.co/openai/whisper-large-v2} to benchmark our systems against competitive state-of-the-art model.

CRDNN combines two blocks of CNN (each block with 2 CNN layers with a channel size (128, 256)), an RNN block (4 bidirectional LSTM layers with 1024 neurons in each layer), and a dense-neural network layer. The inputs are 40-dimensional mel-fiterbank features and the network is trained with an AdaDelta \cite{adadelta} optimizer with a learning rate (LR) of 1 (during fine-tuning we use LR 0.1). The model is trained for 25 epochs with a batch size of 8. During testing, beam search is used with a beam size of 80. Each epoch takes approximately 8h on a single RTXA6000 GPU with 48GB of memory. For wav2vec2.0 and WavLM CTC, training is performed for 45 and 20 epochs respectively with LR 1e-4 on a batch size 8 using an Adam \cite{adam} optimizer. Each epoch takes approximately 5.5h on a single RTXA6000 GPU with 48GB of memory. LR is annealed and the sampling frequency is set to 16 kHz for both approaches. More details on training and model parameters can be found in the repository.

\subsection{Speech enhancement training}
\label{ssec:speech_enhancement}
In this work, we perform speech enhancement using SepFormer \cite{sepformer}-- a multi-head attention transformer-based source separation architecture. It uses a fully learnable masking-based architecture composed of an encoder, a masking network, and a decoder. The encoder and decoder blocks are essentially convolutional layers and we learn a deep-masking network based on self-attention which estimates element-wise masks. These masks are used by the decoder to reconstruct the enhanced signal in the time-domain. We use the DNS4 \footnote[7]{https://github.com/microsoft/DNS-Challenge} dataset to synthesize the training and evaluation set. Using provided clean utterances, noisy clips (150 noises types), and RIRs, we generate 1300h of train and 6.7h of valid set at varying SNR (from -5 dB to 15 dB with a step of 1 dB), and a DNS-2022 baseline dev set is used as test set. Sampling rate is set to 16 kHz and only 30\% of clean speech is convolved with RIR.

SepFormer employs an encoder and decoder with 256 convolution filters with kernel size 16, each with stride 8. The masking network has 2 layers of dual-composition block and a chunk length of 250. With each clean-noisy pairs fixed at 4s in length, the model is trained in a supervised fashion using scale-invariant SNR (SI-SNR) loss and Adam optimizer with LR of 1.5e-4. We utilize multi-GPU distributed data parallel (DDP) training scheme to train the network for 50 epochs with a batch size of 4. Each epoch takes approximately 9h on 8$\times$ RTXA6000 GPU.

\subsection{Training strategies}
\label{ssec:training_methods}

We use various training methods to create a robust speech recognition system that operates in the SAR (Search and Rescue) domain. These methods are described below:
\begin{enumerate}
\item \textit{Clean training}: After pretraining the ASR and Language Model (LM) models, we fine-tune them on the RescueSpeech clean dataset. This process helps to adapt the models to our target domain. We keep the model and training parameters the same as described in Section \ref{ssec:asr}.
\item \textit{Multi-condition training}: Using the same pretrained model as above, we perform multi-condition training, which involves training the ASR model on an equal mix of clean and noisy audio from the RescueSpeech noisy dataset. By doing this, the model can learn to adapt to different noises present in the utterances, which helps it to perform speech recognition. This method forms the baseline for all our results. We set the learning rate (LR) to 0.1 and keep other parameters the same as above.
\item \textit{Model-combination I: Independent training}: We pretrain a speech enhancement model and then fine-tune it on the RescueSpeech noisy dataset. This model is then integrated with the ASR model trained in the \textit{clean training} stage to perform noise-robust speech recognition. In this stage, we freeze the enhancement model.
\item \textit{Model-combination II: Joint training}: This is a continuation of the previous stage, where we follow a joint-training approach. We unfreeze the enhancement model and allow gradients from the ASR to propagate back to the speech enhancement model. Updating the weights of the model in this way enables it to generate output that is as clean as possible, as required by the ASR model.
\end{enumerate}

\section{Results}
\label{sec:results}
\begin{table}[t!]
  \caption{Comparison of test WERs for CRDNN, wav2vec2.0-large, WavLM-large, and whisper-large-v2 models using different training strategies on clean and noisy speech inputs from the RescueSpeech dataset.}
  \label{tab:wer}
  \small
  \centering
  \begin{tabular}[th]{l  c  c |  c}
    \toprule
                                    &   \textbf{ASR Model}         &   \textbf{clean}       &   \textbf{noisy}   \\
    \midrule
    \multirow{2}{*}{Pre-training}   &   CRDNN       &   52.03    &   81.14   \\
                                    &   Wav2vec2    &   47.92    &   76.98   \\
                                    &   WavLM       &   46.28    &   73.84   \\
                                    &   Whisper     &   27.01    &   50.85   \\
    \midrule
    \multirow{3}{*}{Clean training}     &   CRDNN           &   31.18   &   60.10   \\
                                        &   Wav2vec2        &   27.69   &   62.60   \\
                                        &   WavLM           &   23.93   &   58.28   \\
                                        &   Whisper         &   \textbf{23.14}   &   46.70   \\
    \midrule
    \multirow{3}{*}{Multi-cond. training}   &   CRDNN           &   33.22   &   58.95   \\
                                            &   Wav2vec2        &   29.89   &   57.98   \\
                                            &   WavLM           &   25.22   &   52.75   \\
                                            &   Whisper         &   24.11   &   \textbf{45.84}   \\

    \bottomrule
  \end{tabular}
\end{table}

\subsection{ASR Performance}
As a first attempt, we created a simple pipeline consisting solely of an ASR model, with no speech enhancement utilized in the front-end.
Table \ref{tab:wer} provides a comparison of different ASR models used on both clean and noisy audio recordings from the RescueSpeech dataset. The models included in the comparison are CRDNN, wav2vec2.0, WavLM, and Whisper. During the pre-training stage, all models (except Whisper) utilized only the CommonVoice dataset. However, during the clean training and multi-condition fine-tuning stage, the RescueSpeech dataset was used.

Unsurprisingly, the clean training approach is the most effective when tested on clean audio recordings. The top-performing model in this scenario is Whisper, which achieved a WER of 23.14\%. On the other hand, multi-condition training proved to be a superior strategy when dealing with noisy recordings. In this scenario, the best model is again Whisper, which achieved a WER of 45.84\%. The performance gap with clean signals, highlights one more time the significant decline in recognition performance when dealing with challenging acoustic conditions, even for models that were pre-trained using state-of-the-art self-supervised techniques like wav2vec, WavLM, and Whisper (the latter of which is even semi-supervised).

\begin{table}[t!]
  \caption{Speech enhancement performance on the RescueSpeech noisy test inputs when combining speech enhancement and speech recognition (Model Comb. I vs Model Comb. II).}
  \label{tab:model_combination}
  \centering
  \footnotesize
  \begin{tabularx}{\columnwidth}{l|X|XXXX}
    \toprule
                    & \multirow{2}{*}{\shortstack{\textbf{Model} \\ \textbf{Comb. I}}} & \multicolumn{4}{c}{\textbf{Model Comb. II}} \\
                    \cmidrule{3-6}
                    &       &   \scriptsize{CRDNN}   &   \scriptsize{wav2vec2}  &   \scriptsize{WavLM}   &   \scriptsize{Whisper} \\
                    \midrule
            SI-SNRi	    &   6.516	& 6.618       &   7.205    &   7.140     &   7.482      \\
            SDRi        &   7.439	& 7.490	      &   7.765    &   7.694     &   8.011       \\
            PESQ        &   2.008	& 2.010       &   2.060    &   2.064	 &   2.083       \\
            STOI        &   0.842	& 0.844       &   0.854    &   0.854	 &   0.859         \\

    \bottomrule
  \end{tabularx}
\end{table}

\begin{table}[t!]
  \centering
  \footnotesize
  \begin{center}
    \caption{Word-Error-Rate (WER\%) achieved with independent training (Model Comb. I ) and joint training (Model Comb. II) of the speech enhancement and ASR modules.}
    \label{tab:model_combination2}
    \begin{tabular}{l|c|c} 
    \toprule
      \textbf{ASR Model} & \textbf{Model Comb. I} & \textbf{Model Comb. II} \\
      \midrule
      CRDNN     &   54.98   &   54.55\\
      Wav2vec2  &   50.68   &   49.24\\
      WavLM     &   48.24   &   46.04\\
      Whisper   &   48.04   &   \textbf{45.29}\\
       \bottomrule
    \end{tabular}
  \end{center}
\end{table}

\begin{figure}[t!]
  \centering
  \includegraphics[scale=0.50]{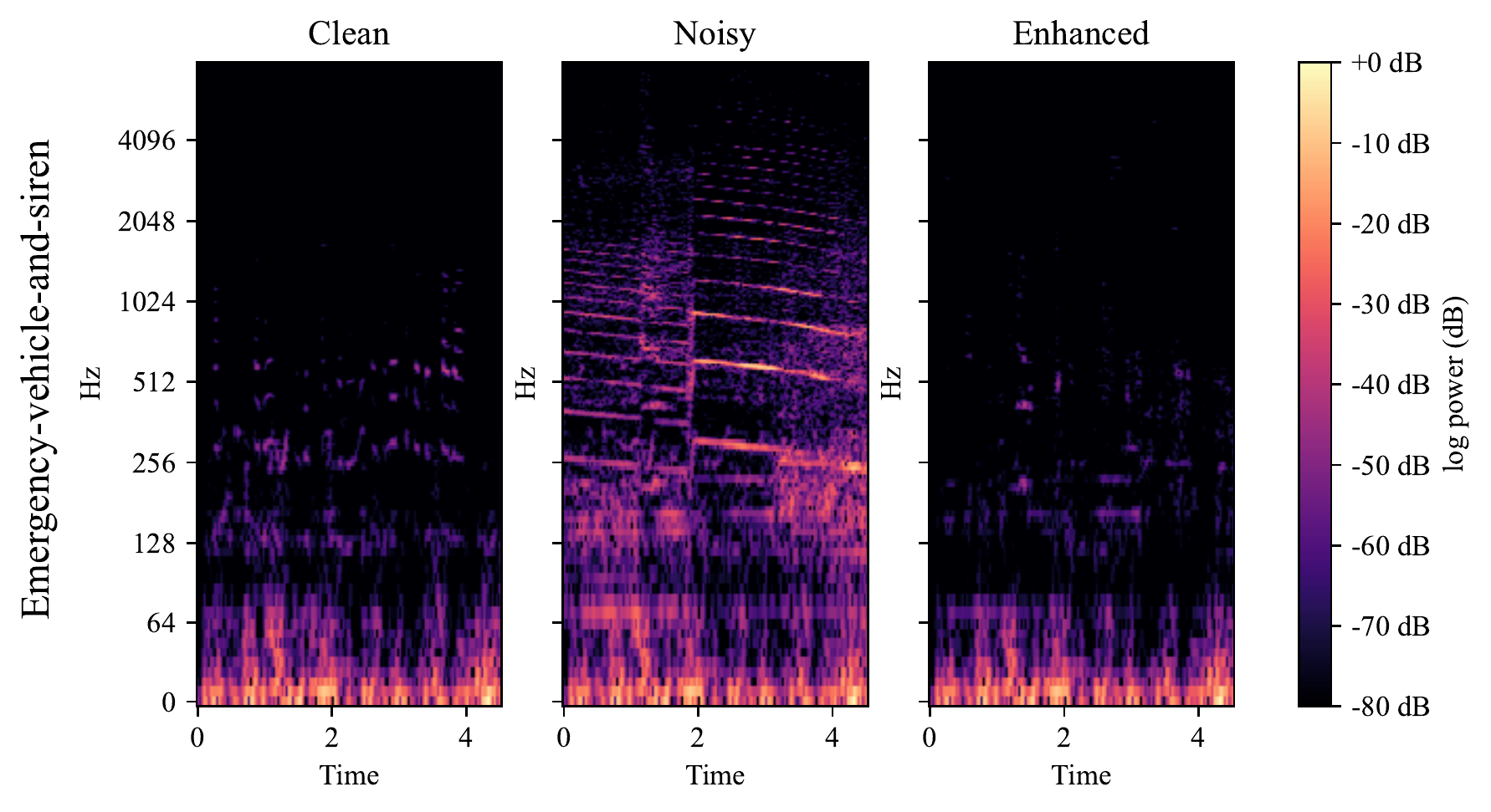}
  \includegraphics[scale=0.50]{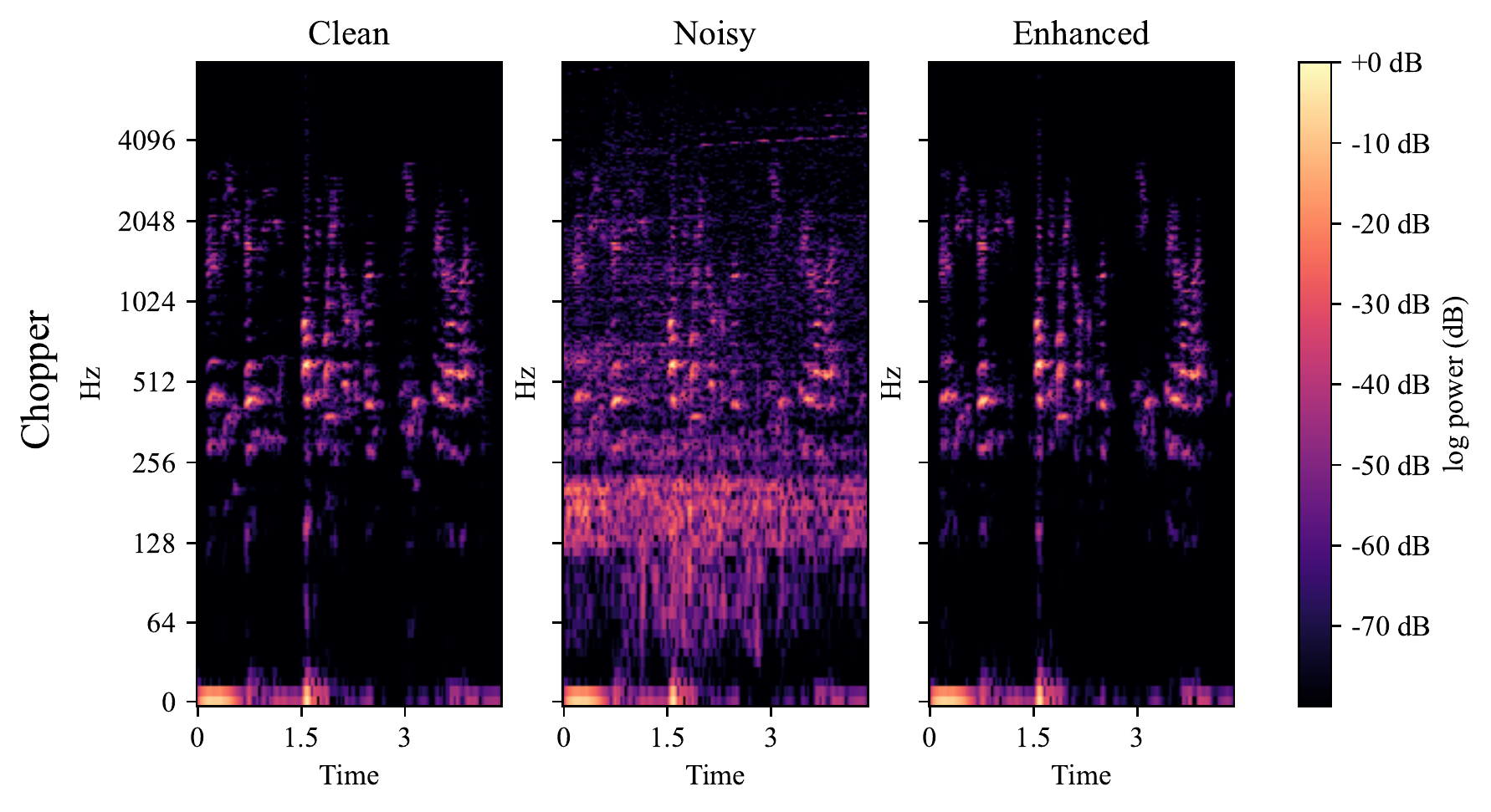}
  \caption{Log-power spectrogram of clean, noisy, and SepFormer-enhanced utterances for \textit{emergency vehicle siren} and \textit{chopper} noise types at -5 dB SNR.}
  \label{fig:spectrogram}
\end{figure}

\subsection{Combining ASR and Speech Enhancement}
\begin{figure}[t!]
  \centering
  \includegraphics[scale=0.4]{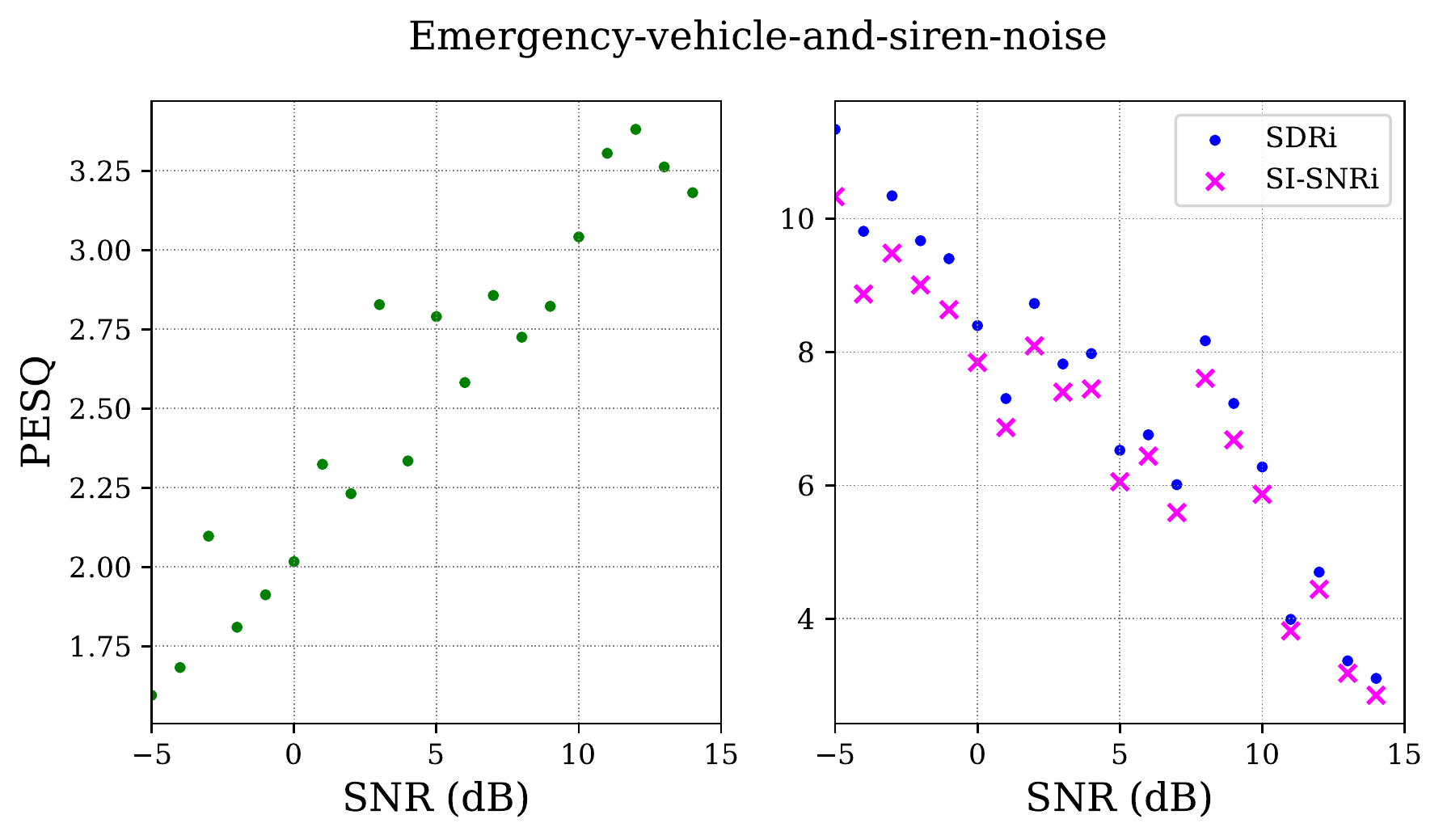}
  \includegraphics[scale=0.4]{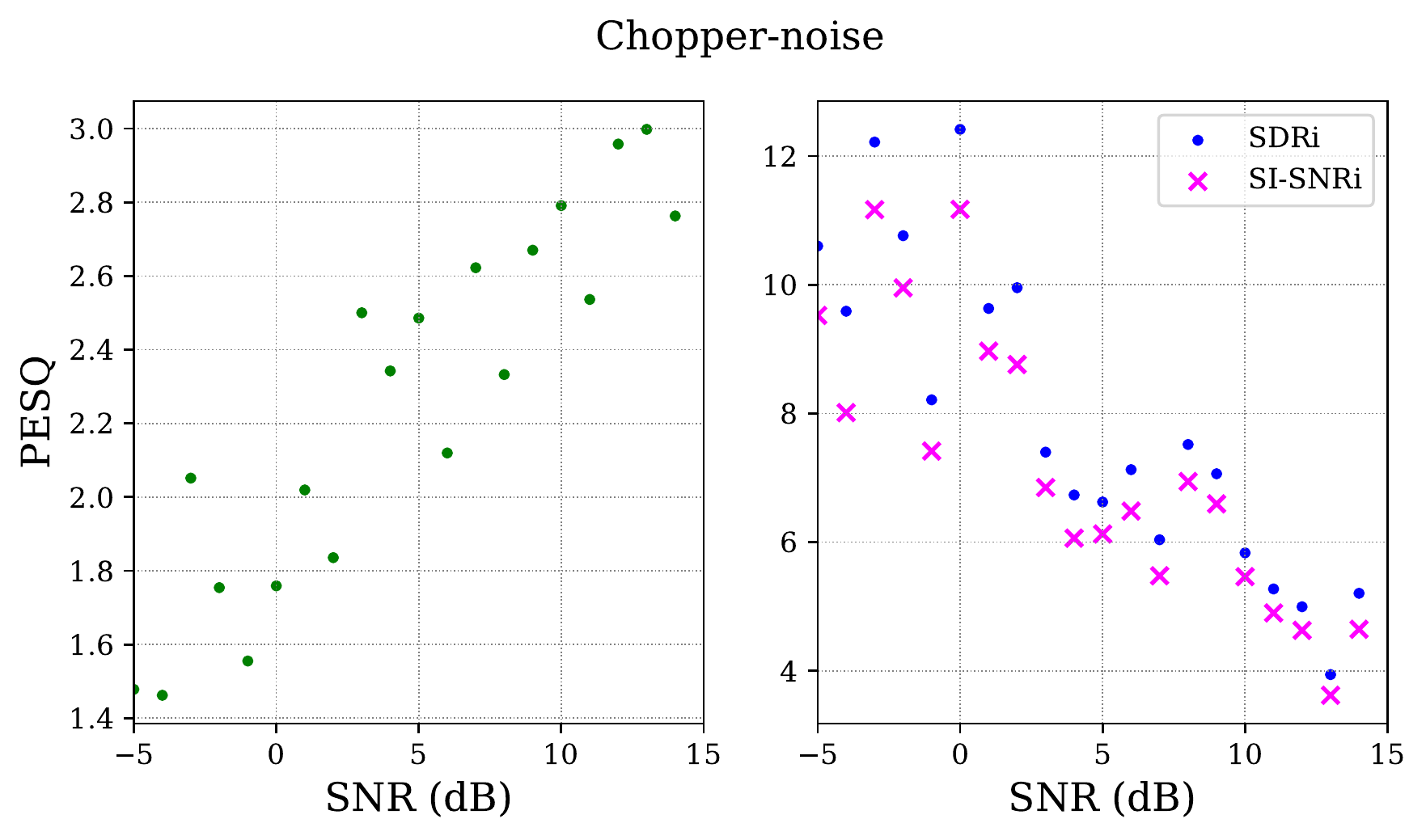}
  \caption{PESQ, SDRi, SI-SNRi vs SNR of SepFormer enhanced utterances for two noise types-- \textit{emergency vehicle siren} and \textit{chopper} noise.  
  }
  \label{fig:metric_plot}
\end{figure}

In order to improve the ASR performance, we developed a speech enhancement system to clean up the recordings. To accomplish this, we utilized the SepFormer model, which has demonstrated competitive performance in speech separation and enhancement tasks \cite{sepformer2}. Specifically, we trained the model on the DNS4 dataset, achieving SIG, BAK, and OVRL scores of 2.999, 3.076, and 2.437, respectively.
Figure \ref{fig:spectrogram} shows the log-power spectrogram for two types of noisy audio recordings, \textit{emergency vehicle siren} and \textit{chopper noise}, both with an SNR of -5 dB, using the SepFormer model fine-tuned on the RescueSpeech noisy dataset. From a qualitative standpoint, it appears that SepFormer performs well on noises that impact the SAR domain.
Figure \ref{fig:metric_plot} presents PESQ vs SNR and SI-SNRi, SDRi vs SNR for the same noise types. We observed that improvements in SI-SNR and SDR were greater for utterances with an SNR of -5 dB, indicating a more significant enhancement in speech intelligibility and reduction of distortion than for higher SNR utterances. This pattern is consistent across all noise types.

Table \ref{tab:model_combination} displays the speech enhancement results obtained by incorporating a speech recognizer into the pipeline. In section 3.3, we explored two approaches: independent training (Model Comb. I) and joint training (Model Comb. II). The joint training approach resulted in improvements across all considered speech enhancement metrics (SI-SNRi, SDRi, PESQ, STOI) and all ASR modules (CRDNN, Wav2vec2, WavLM, Whisper).
Table \ref{tab:model_combination2} presents the final speech recognition output at the end of the pipeline.

As anticipated, the joint training approach outperformed a simple combination of independently trained speech enhancement and speech recognition modules. It is important to note that both speech enhancement and speech recognition models undergo fine-tuning using enhanced signals from the unfrozen Sepformer. We postulate that backpropagating the ASR gradient to the speech enhancement model enables the SepFormer to denoise utterances according to the specific requirements of the ASR model, facilitating better convergence. Training both models jointly allows the enhancement model to adapt its cleaning capabilities to align better with the needs of the ASR system. Overall, the best-performing model is the combination of SepFormer with Whisper ASR, which achieved a WER of 45.29\%.

\section{Conclusions}
\label{sec:conclusions}
Our work addresses some major challenges that arise in the SAR domain: the lack of speech data, the need for robustness to SAR noises, and conversational speech. To overcome these challenges, we have introduced RescueSpeech, a new dataset of speech data in German that we use to perform robust speech recognition in a hostile noise-filled environment.
To achieve this, we proposed multiple training strategies that involve fine-tuning pretrained models on our in-domain data. We tested different self-supervised models (e.g, Wav2Vec2, WavLM, and Whisper) for speech recognition. Despite leveraging these cutting-edge systems, our best model only achieves a WER of 45.29\% on our test set. This result highlights the significant difficulty and the urgent need for further research in this crucial domain.

Overall, our work represents a step forward in addressing the challenges of speech recognition in the SAR domain. By introducing a new dataset, we hope to establish a useful benchmark and foster more studies in this field.

\section{Acknowledgements}
\label{sec:acknowledgments}
Our work was supported under the project ``A-DRZ: Setting up the German Rescue Robotics Center'' and funded by the German Ministry of Education and Research (BMBF), grant No. I3N14856. We would like to thank our colleague from A-DRZ project- Alina Leippert for transcribing the dataset.






\bibliographystyle{IEEEbib}
\bibliography{strings,refs}

\end{document}